# Is Voyager 1 Inside an Interstellar Flux Transfer Event?

N.A. Schwadron[1,2*] and D.J. McComas[2,3]

[1] University of New Hampshire, Durham, NH, USA

[2] Southwest Research Institute, San Antonio, Texas, USA

[3] University of Texas at San Antonio, San Antonio, TX, USA

*Correspondence to: n.schwadron@unh.edu

**Abstract.** Plasma wave observations from Voyager 1 have recently shown large increases in plasma density, to about $0.1$ cm$^{-3}$, consistent with the density of the local interstellar medium. However, corresponding magnetic field observations continue to show the spiral magnetic field direction observed throughout the inner heliosheath. These apparently contradictory observations may be reconciled if Voyager 1 is inside an interstellar flux transfer event – similar to flux transfer events routinely seen at the Earth's magnetopause. If this were the case, Voyager 1 remains inside the heliopause and based on the Voyager 1 observations we can determine the polarity of the interstellar magnetic field for the first time.

**Keywords:** Sun: heliosphere; ISM: general; (ISM:) cosmic rays; ISM: magnetic fields; (Galaxy:) local interstellar matter

**1. Introduction.** The Voyager 1 and 2 (V1, V2) spacecraft continue to move outward from the Sun, venturing deep into the outermost reaches of our solar system and heliosphere. The boundary separating the solar wind from the interstellar plasma is called the heliopause and is expected to be marked by a large increase in plasma density, from ~0.002 cm$^{-3}$ in the inner heliosheath inside the heliopause, to ~0.1 cm$^{-3}$ in the interstellar medium. The heliopause separates the solar wind created by the Sun from the ionized portion of the interstellar medium. On April 9, 2013, V1's plasma wave instrument detected electron plasma oscillations at a frequency of ~2.6 kHz. These oscillations are created locally with frequency corresponding electron density ~0.08 cm$^{-3}$, orders of magnitude larger than in the inner heliosheath. These observations provide evidence that V1 had crossed the heliopause (Gurnett et al. 2013).

The first observations suggesting that V1 may have encountered the heliopause were made on July 28, 2012, when V1 was at 121 AU and the Low Energy Charged Particle (LECP) and Cosmic Ray (CRS) instruments showed abrupt decreases in the fluxes of energetic particles (EPs, termed "termination shock particles" or TSPs) and anomalous cosmic rays (ACRs) (Krimigis et al. 2013, Stone et al. 2013, Webber & McDonald 2013). These decreases in particles energized by the termination shock (Pesses et al. 1981, Zank 1999, McComas & Schwadron 2006, Schwadron & McComas, 2007, Schwadron et al. 2008, Kóta & Jokipii 2007, McComas & Schwadron 2012) were correlated with increases in the galactic cosmic ray (GCR) intensity (Stone et al. 2013, Webber & McDonald 2013). Five similar crossings of the boundary were observed between July 28, 2012 and August 25, 2012. The decrease in EPs and ACRs and coincident increase in GCRs strongly supports the conclusion that Voyager 1 was either moving into the interstellar medium during boundary crossings or that the spacecraft was moving into regions magnetically connected to the interstellar medium.

The V1 magnetometer (MAG) observations (e.g., Burlaga et al. 2005, Burlaga et al. 2013) have confounded the interpretation that V1's boundary crossings were simply heliopause crossings. Although boundary crossings were correlated with changes in the magnetic field strength, the magnetic field direction remained extremely constant and consistent with the spiral magnetic structure observed throughout the inner heliosheath. In contrast, the magnetic field in the local interstellar medium (Lallement et al. 2005, Opher et al. 2006, Pogorelov et al. 2009a, 2009b, Frisch et al. 2012, Frisch & Schwadron 2013) is expected to be quite different from the spiral magnetic field of the inner heliosheath. During the detection of increased plasma density at V1 on April 9, 2013, the magnetic field apparently remained at least roughly in the direction of the spiral magnetic field associated with the inner heliosheath (Gurnett et al. 2013).

**2. Interstellar Flux Transfer Events.** We suggest here that instead of crossings of the heliopause, V1's boundary crossings might instead have been into Interstellar Flux Transfer Events (IFTEs; Figure 1) still inside the heliopause, akin to a phenomenon – flux transfer events (FTEs) – routinely observed at Earth's magnetopause (e.g., Russell & Elphic 1979, Fuselier et al. 2011, Tkachenko et al. 2011, Zhang et al. 2011, Fear et al. 2012).

The configuration of the heliosheath is in many respects similar to that of the magnetosphere. The heliopause, like the magnetopause, separates distinct plasma regions. And like the magnetosphere relationship to the solar wind, the state of the heliosheath is firmly linked to the conditions in the interstellar medium. In the magnetosphere, when the interplanetary magnetic field (IMF) has a northward orientation, parallel to the magnetospheric field, there is very little energy transfer from the solar wind resulting in quiet geomagnetic conditions. In contrast, when the IMF has a southward orientation, the solar wind transfers greater energy into the magnetosphere, which leads to increased geomagnetic activity. Dungey (1961) suggested that the mechanism controlling the solar wind energy transfer from the solar wind into the magnetosphere is magnetic reconnection. The first direct in-situ evidence for quasi-steady state magnetic reconnection came from plasma and magnetic field measurements from ISEE 1 and 2 (Paschmann et al. 1979, Sonnerup et al 1981, Gosling et al, 1982). However, even when solar wind conditions are favorable for reconnection, it is not always present at magnetopause crossings. This demonstrates that magnetic reconnection is fundamentally sporadic in time and patchy in space (Haerendel et al. 1978). Russell & Elphic (1978) found isolated structures in the magnetic field and plasma of the low-latitude magnetosheath. They appeared to be passages through localized, reconnected flux tubes, and so were dubbed flux transfer events (FTEs).

Figure 1 shows the global configuration of the heliopause, where based on the analogy with magnetospheric reconnection, we postulate the existence of bursty intervals of reconnection with the interstellar magnetic field. In this case, the interstellar magnetic field should be quite steady over years. If the interstellar magnetic field has a significant component along the heliopause nose, then magnetic reconnection will be preferred when the spiral field inside the heliopause opposes the interstellar magnetic field. The observations of Burlaga et al. (2013) demonstrate that the intervals where we observe rapid reductions in ACR fluxes and coincident increases in the magnetic field magnitude (from ~0.2 nT to ~0.4 nT) occur where the magnetic field is oriented in the starboard[1] direction of the heliosphere. Therefore, the interstellar magnetic field should have a port-directed component.

Analogously to FTEs at Earth, IFTEs created from patchy reconnection would be pulled into the inner heliosheath as the magnetic tension force attempts to relax (or straighten) the field. The key advantage of an IFTE for explaining the Voyager 1 observations is that it creates a magnetic connection out through the heliopause, providing a pathway to drain heliospheric EPs and access to GCRs while still maintaining the heliospheric magnetic field orientation. Therefore, an IFTE would be expected to be associated with a strong increase in density as observed by Voyager 1's plasma wave instrument in addition to an abrupt reduction in the fluxes EPs and ACRs as these particles stream out along the flux tube beyond the heliopause. This same magnetic connection provides ready access for the entry of GCRs explaining the large enhancements observed in GCR fluxes at the boundary crossings. Notably, these EP and plasma signatures would be accompanied by little change in the direction of the magnetic field since inside the heliopause an IFTE flux tube maintains the magnetic structure of the inner heliosheath. However, the loss of EPs and ACRs on the magnetic flux tube of the IFTE constitutes a substantial reduction in plasma pressure. Since the magnetic flux tube remains in the inner heliosheath, it must attain pressure balance through magnetic compression and subsequent increase in the flux tube's field strength as observed (Burlaga et al. 2013).

---

[1] Port and startboard are nautical terms applied to the heliosphere by McComas et al. 2013. The Port direction is shown in Figure 1.

114    Within the magnetosphere, FTEs have sizes of ~1 $R_e$, last typically for ~1 minute, occur every 8 minutes, and are observed over rough half of the magnetopause at any given time (e.g., Russell 1995, Haerendel et al. 1978, Russell & Elphic, 1978; Berchem & Russell 1984). Russell (1995) studied FTEs at the planets and found that they change in size, duration and frequency roughly in proportion to the size of the magnetosphere. In the case of the heliopause, we estimate the inner heliosheath thickness near the nose as ~40 AU, implying a ~4 AU size-scale for IFTEs, ~2 month duration and ~1.4 year frequency. The Alfven speed near the heliopause (~40 km/s) represents the characteristic propagation speed of IFTEs away from the point of reconnection. Over the ~2 month duration of an IFTE, it propagates ~1.5 AU, which is comparable to the estimated size of IFTEs. While this demonstrates some internal consistency in the size and duration of IFTEs, these estimates are preliminary. Further, as for FTEs, we expect significant variability in IFTEs.

127    Paschmann et al. (1982) studied the plasma properties of FTEs, which prompted Thomsen et al. (1987) to analyze FTE plasma ion and electron phase space distributions that consist of a mixture of magnetospheric and magnetosheath populations. Thomsen et al (1987) found a depletion of the higher EPs relative to the magnetospheric distribution indicating that magnetic reconnection facilitates the leakage of higher EPs out of the magnetosphere. The depletion is greatest at the center of the FTE and the distributions of EPs show significant variability within FTEs.

134    Similar to FTEs, IFTEs provide diffusive leakage of ACRs out of the inner heliosheath and access of GCRs from beyond the heliopause into the inner heliosheath. Schwadron and McComas (2007) estimate a parallel scattering mean free path of $\lambda_{\parallel} = \lambda_0$ 137 $(R/R_0)^{1/2}$ where $\lambda_0$ =50 AU, $R$ is the EP rigidity, and $R_0$=1.5 GV. Over the $\tau$ ~2 month

138 duration of an IFTE, EPs diffusively propagate over a characteristic distance of $d=(2\tau\kappa_{||})^{1/2}$

139 where the parallel diffusion coefficient is $\kappa_{||} = \lambda_{||} v/3$ and $v$ the particle speed. At 1

140 MeV (ACR protons), this diffusive length is $d$~50 AU and at 100 MeV (GCR protons)

141 the diffusive length is 300 AU, demonstrating that ACRs leak out of the inner heliosheath

142 rapidly and GCRs access large regions within the heliopause after magnetic reconnection.

143 Another factor is diffusion across the magnetic field. Consider a ratio of

144 perpendicular to parallel diffusion $\kappa_\perp/\kappa_{||}$ ~1% (Giacalone & Jokipii, 1999), which implies

145 a cross-field mean free paths ~0.08 AU at 1 MeV and ~0.3 AU at 100 MeV. These

146 perpendicular mean free paths are smaller than the scale of the IFTE indicating that the

147 IFTE can entrain EPs at ACR and GCR energies. If we assume the IFTE moves at the

148 Alfven speed of ~40 km/s over an observer (e.g., the Voyager 1 satellite), the time it

149 takes to move across the perpendicular mean free path is ~4 days at 1 MeV and ~12 days

150 at 100 MeV, thereby constituting very sharp gradients. However, diffusion acts to

151 broaden energetic particle gradients. Consider the perpendicular diffusion scale, $d_\perp$~$(2 t_d$

152 $\kappa_\perp)^{1/2}$ ~ $l (\kappa_\perp/\kappa_{||})^{1/2}$ where $l$ is the distance along the field to the reconnection location,

153 $\kappa_\perp$ is the perpendicular diffusion coefficient, and $t_d = l^2/(2\kappa_{||})$ is the diffusion time to the

154 observer along the field. At low energies, the perpendicular diffusion scale may exceed

155 the mean free path (e.g., $l$~10 AU results in $d_\perp$~0.5 AU) and thereby broaden the

156 observed energetic particle gradient, particularly at ACR energies. The sharpness of

157 observed energetic particle gradients therefore limits the possible distance from Voyager

158 1 to the reconnection point. If an IFTE typical size is ~4 AU and moves ~40 km/s over an

159 observer, it takes roughly 6 months for the structure to sweep over the observer. Over this

160 period, V1 would observe different structures within the IFTE associated with mixing of

161 heliosheath and interstellar plasmas as well as varying magnetic field strengths and the

162 appearance of EP boundaries associated with magnetic connectivity to the heliosheath

163 and interstellar medium. This same complexity observed in FTEs is attributed to the non-

164 steady nature of bursty reconnection (Haerendel et al. 1978, Russell & Elphic 1978).

165 Table 1 summarizes the analogy between FTEs and IFTEs. The component

166 directions of the magnetic field are shown in Figure 1. One of the key observational

167 diagnostics for FTEs is a $B_n$ component with a sign that reverses from one side of the FTE

168 to the other. This change in the normal component arises from the FTE flux rope

169 configuration. The magnitude of this normal component increases toward the middle of

170 the FTE. In contrast to FTEs, the large-size of IFTEs makes it difficult to observe

171 significant portions of the structure over brief periods of time. On the outskirts of the

172 IFTE, the variation in the elevation angle and azimuthal angle of the IFTE is expected to

173 be small. However, these variations intensify closer to the center of the IFTE.

174 The magnetic reconnection needed to produce IFTEs can only occur for the

175 portions of the inner heliosheath where the magnetic field has an orientation with a

176 component anti-parallel to interstellar magnetic field at the heliopause. This constraint

177 implies that one magnetic sector of the inner heliosheath (the outward sector in Figure 2

178 where V1 currently resides) will tend to undergo magnetic reconnection. An important

179 implication is that the observation of IFTEs by V1 should only be associated with this

180 one of the two opposite sector orientations.

181 One of the long-standing questions of heliospheric physics is the direction and

182 strength of the local interstellar magnetic field. If the IFTE picture correct then the outward magnetic sector must go through a greater than 90° rotation to connect to the interstellar magnetic field direction. The field direction has been estimated previously based on Lyman-alpha observations (Lallement et al. 2005), using large-scale models (Opher et al. 2006, Pogorelov et al. 2009a, 2009b) and from the center of the IBEX ribbon (McComas et al. 2009, Schwadron et al. 2009, Funsten et al 2013), which yields a magnetic field direction (λ, β) ~ (220°, 38°) or ~ (40°, -38°) at 2-3 keV in ecliptic coordinates. The 180° ambiguity in the interstellar field vector was here-to-for unsolvable with existing measurements.

The longitude of the ribbon center is *smaller* than that of the heliosphere's nose (259°; McComas et al. 2012, Bzowski et al. 2012, Mobius et al. 2012). For magnetic reconnection at the heliopause, the longitudinal draping of the interstellar field along the heliopause should oppose the azimuthal direction of the outward sector magnetic field. V1 (λ=255.3°, β=35°) is near the nose in longitude so the outward sector magnetic field extends from the port to starboard direction. Therefore, we expect the interstellar field transverse component to extend from starboard to port, which resolves the 180° ambiguity in the interstellar field vector and defines its *polarity*. The interstellar magnetic field direction associated with the IBEX ribbon center is (λ, β) ~ (40°, -38°) in ecliptic coordinates.

**3. Summary.** Table 2 compares how proposed concepts address the key observables at V1 boundary crossings. These concepts are summarized here:

- **The zero-radial-speed boundary**: The passage of V1 and V2 through the inner heliosheath has revealed a region where radial solar wind speed slows down from ~400 to ~125 km/s (Richardson et al. 2008) and may decrease to very low values even further from the termination shock (Krimigis et al. 2011). Krimigis et al. (2011) suggest that Voyager 1 has entered a special transition layer of zero-radial-velocity plasma flow bounded by the inner heliosheath and the interstellar plasma. It is currently unclear what drives the zero-radial-velocity. The boundary region is not connected to the termination shock, which implies that ACRs and EPs should be depleted. Further, depending on the global configuration of the boundary and its possible connection across the heliopause, it may provide preferred access to GCRs. Suprathermal particles should play a significant role in determining the dynamics of the boundary layer. However, the boundary region is distinct from the ISM and should not provide preferred access to ISM plasma.

- **The porous boundary:** particle-in-cell simulations by Swisdak et al. (2013) suggest that the sectored region of the heliosheath through which V1 travelled may produce large-scale magnetic islands that reconnect with the interstellar magnetic field while mixing local interstellar and inner heliosheath particles. Therefore, V1's detection of a higher density plasma may suggest that the heliopause is a porous boundary where magnetic reconnection creates complex magnetic structures. Because of the magnetic complexity, we do not expect the field direction to match that of the inner heliosheath, nor do we expect plasma changes to be uniquely associated with sector crossings. Because the porous boundary is distinct from the inner heliosheath and magnetically connected to the ISM, we expect correlation of ACR and EP depletions with GCR enhancements.

- **The Disconnection Boundary:** McComas & Schwadron (2012) showed that reductions in ACRs, EPs and increases in GCRs naturally arise from the heliosphere's global magnetic topology. For a blunt termination shock (McComas & Schwadron, 2006), there is a region of magnetic flux inside the heliopause that represents the last magnetic connection point to the termination shock. Beyond this disconnection boundary, there is poorer access to the shock-accelerated anomalous cosmic rays and better access for the galactic cosmic rays entering the heliosphere. However, the disconnection boundary itself provides no means to transfer plasma from the interstellar medium into the inner heliosheath. Further, we would not expect dropouts in ACRs and enhancements in GCRs to be associated with sector crossings.

- **Heliopause Crossing:** Passage into the interstellar medium through the heliopause would naturally lead to reductions in ACRs and EPs produced inside the heliopause. GCR intensities would be enhanced to their high levels in the interstellar medium. The difficulty is that we do not expect the magnetic structure of the inner heliosheath to persist beyond the heliopause. It follows that we should not expect changes at sector crossings. Quantifying this point, Burlaga et al. (2013) reports that the field changes direction by less than 2° at V1 boundary crossings. If the interstellar magnetic field had an equal probability of being oriented in any direction, the probability that it would be oriented in a direction within 2° of that inside the heliopause (regardless of polarity) is 0.06%, which is highly unlikely.

- **IFTE:** IFTEs provide magnetic access to the interstellar medium, while maintaining a magnetic structure similar to that of the inner heliosheath. This magnetic connection naturally causes dropouts in ACRs and EPs and

ASTROPHYSICAL JOURNAL LETTERS, In press, 2013enhancements in GCRs. Further, because magnetic reconnection is active only for outward sectors, we expect changes at or near sector crossings. The IFTE concept therefore explains the observations of Voyager 1 at boundary crossings. If the IFTE concept is correct, we expect that V1 will continue to observe boundary crossings, and that as V2 approaches the heliopause it will observe IFTEs for the same outward directed magnetic sectors as observed by V1.

Thus, we suggest that V1 has entered an IFTE inside the heliopause. FTEs are commonly observed just inside the Earth's magnetopause. The evolution of an IFTE provides naturally for the observed changes in plasma, EPs and cosmic rays observed by V1 while retaining the magnetic field direction associated with the inner heliosheath. The IFTE concept along with other hypotheses including passage into interstellar space will be tested by the increasing wealth of observations from both Voyagers.

Acknowledgements. NAS is grateful for discussion with Eberhard Moebius. This work was supported by the Interstellar Boundary Explorer mission of NASA's Explorer Program.

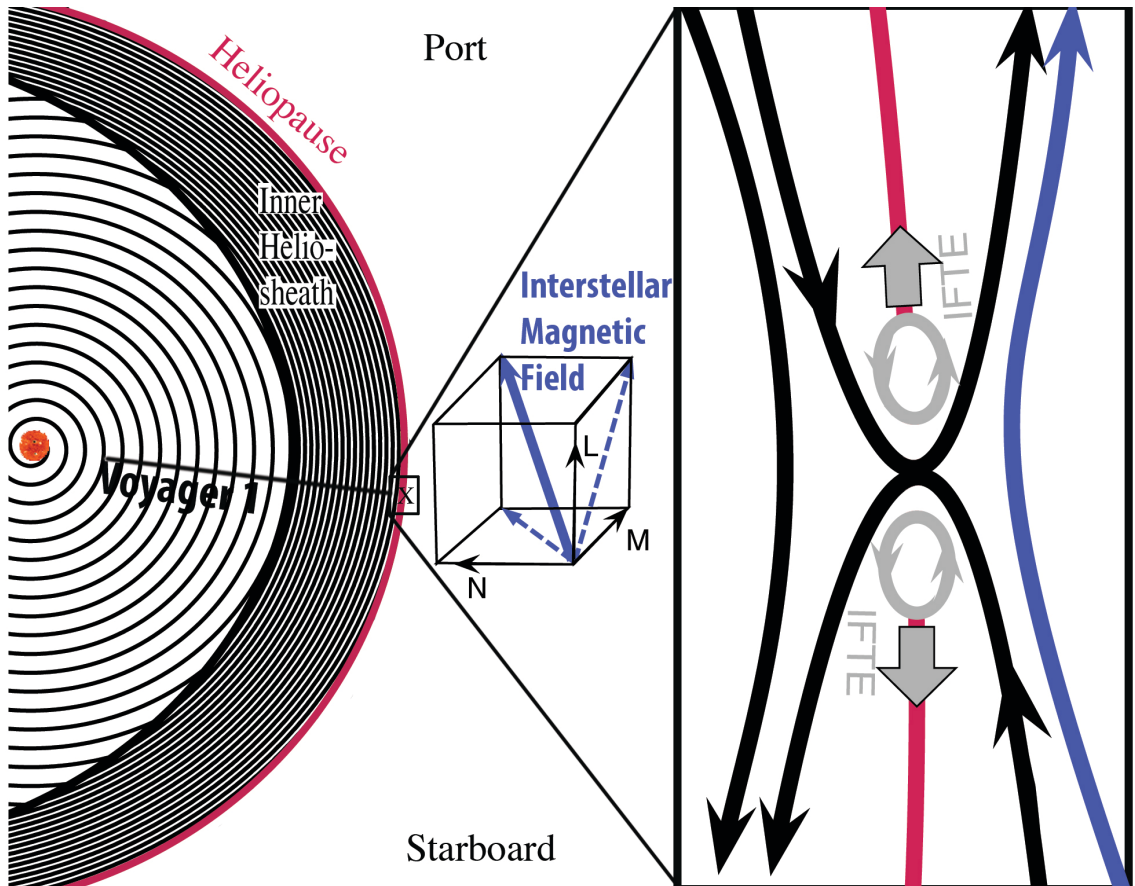

**Figure 1**. Global configuration of the heliopause showing regions where reconnection between the interstellar magnetic field and the magnetic field of the inner heliosheath are conducive to the formation of IFTEs. Note that this configuration depends on achieving a specific polarity of the magnetic field of the inner heliosheath. Since the magnetic field polarity is outward (-L component), this would imply a strong (Port-directed) component of the interstellar magnetic field. The projection here looks down from the north ecliptic pole with L, M, N components in the port-, south-, and radial-inward direction, respectively.

346

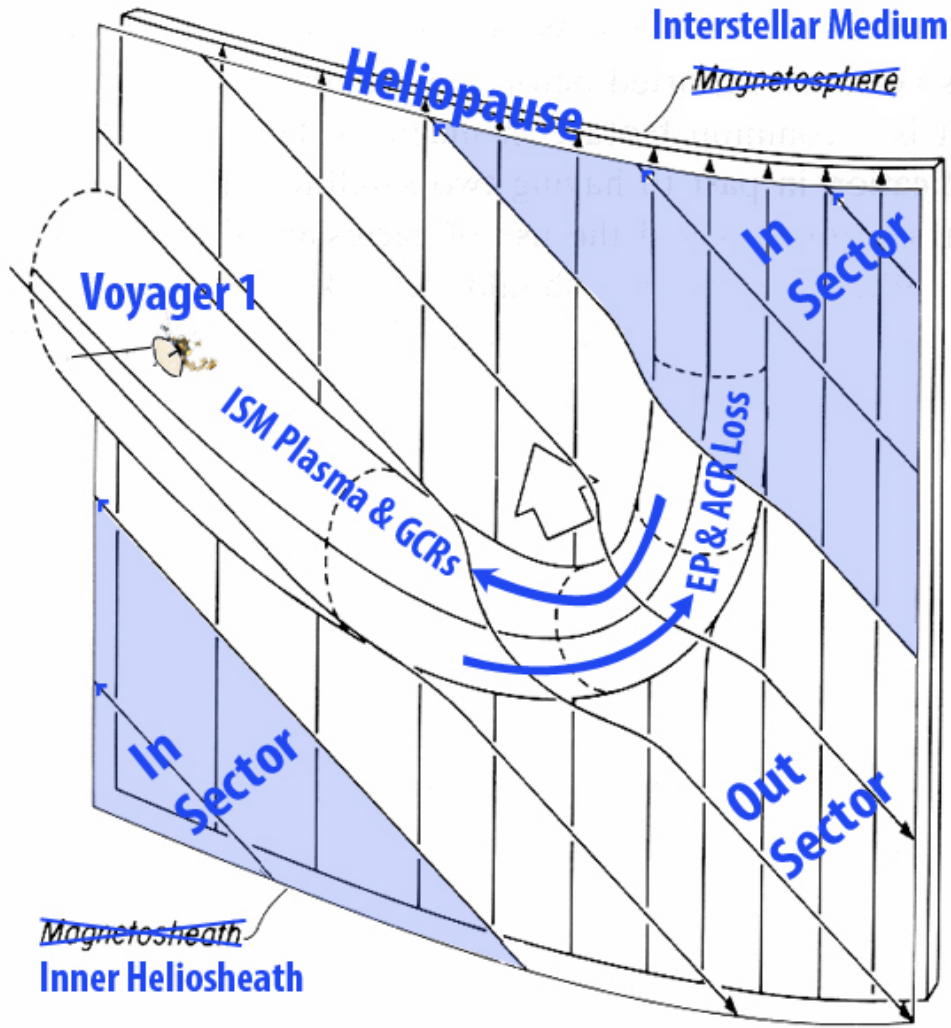

347

**Figure 2.** The IFTE adapted from an illustration of a similar phenomenon in the magnetosphere (Russell & Elphic, 1979). Patchy magnetic reconnection at the heliopause leads to the creation of a magnetic flux tube inside the heliopause but connected out into the local interstellar medium. This magnetic connection allows access for higher density interstellar plasma and GCRs and loss of TSPs and ACRs, while maintaining the dominant spiral magnetic field direction of the inner heliosheath. Note that the reconnection with the magnetic field in the local interstellar medium will be

355    preferred for only one magnetic field sector of the heliosheath. The vertical lines indicate

356    the magnetic field direction of the local interstellar medium. The flux tube shown here

357    bends as it transitions from interstellar magnetic field orientation to outward sector

358    magnetic orientation inside the heliopause.

**Table 1.** Table comparing IFTE and FTE taxonomy.

| Concept | FTE (near magnetopause) | IFTE (near heliopause) |
|---|---|---|
| $B_L$ | Steady<br>Increased strength[1] | Steady (+ inside heliopause)<br>Increased strength |
| $B_N$ | +/- or -/+<br>Magnitude of oscillation depends on depth within FTE[1] | Small deviation in azimuthal angle predicted on outskirts of IFTE. Deviation angle increases closer to the interior of IFTE. Sign of deviation depends on location of V1 with respect to IFTE |
| $B_M$ | Magnitude enhancement[1] | Small deviation in elevation angle on outskirts of IFTE. Deviation angle increases closer to the interior of IFTE. Sign of deviation depends on location of V1 with respect to IFTE. |
| Plasma | Density enhancement (larger enhancements closer to FTE center)[2] | Density enhancement (larger enhancements closer to IFTE center) |
| Energetic Particles | Loss of magnetosphere populations[3] | - Loss of Anomalous Cosmic Rays<br>- Increase in Galactic Cosmic Rays |
| Size | ~1 Re[1] | ~4 AU |
| Frequency | ~ 8 minutes[1] | ~1.4 years |
| Duration | ~1 minute[1] | ~2 months |

1: Russell 1995, Haerendel et al. 1978, Russell & Elphic, 1978; Berchem & Russell 1984
2: Paschmann et al. 1982, Thomsen et al. 1987
3: Thomsen et al 1987

**Table 2.** Truth table comparing models and concepts potentially explaining V1 boundary crossings.

| Concept | ACR & EP Depletion[1] | GCR Increase[2] | Inner Hsh Field direction[3] | ISM Plasma Density[4] | Plasma & EP changes at Sector crossing[3] |
|---|---|---|---|---|---|
| **Zero-radial-speed boundary**[5] | Yes | Possibly | Yes | No | No |
| **Porous Boundary**[6] | Yes | Yes | No | Yes | No |
| **Disconnection Boundary**[7] | Yes | Yes | Yes | No | No |
| **Heliopause Crossing**[4] | Yes | Yes | No | Yes | No |
| **IFTE** | Yes | Yes | Yes | Yes | Yes |

1: Krimigis et al. 2013, Stone et al. 2013.
2: Stone et al. 2013, Webber & McDonald 2013.
3: Burlaga et al. 2013
4: Gurnett et al. 2013
5: Krimigis et al. 2011
6: Swisdak et al. 2013
7: McComas & Schwadron 2012